\documentclass{nature}

\usepackage{graphicx}
\usepackage{lineno}

\title{Quantum spin liquids by geometric lattice design}

\author{Xiaoran Liu$^{1\ast}$, T. Asaba$^2$, Qinghua Zhang$^3$, Yanwei Cao$^1$, B. Pal$^{1,4}$, S. Middey$^5$, P. S. Anil Kumar$^5$, M. Kareev$^1$, Lin Gu$^3$, D. D. Sarma$^4$, P. Shafer$^6$, E. Arenholz$^6$, J. W. Freeland$^7$, Lu Li$^2$ \& J. Chakhalian$^1$}

\begin{document}

\maketitle

\begin{affiliations}
 \item Department of Physics and Astronomy, Rutgers University, Piscataway, New Jersey 08854, USA.
 \item Department of Physics, University of Michigan, Ann Arbor, Michigan 48109, USA.
 \item Beijing National Laboratory for Condensed-Matter Physics and Institute of Physics, Chinese Academy of Sciences, Beijing 100190, P. R. China.
 \item Department of Solid State and Structural Chemistry Unit, Indian Institute of Science, Bengaluru 560012, India.
 \item Department of Physics, Indian Institute of Science, Bengaluru, 560012, India.
 \item Advanced Light Source, Lawrence Berkley National Laboratory, Berkeley, California 94720, USA.
 \item Advanced Photon Source, Argonne National Laboratory, Argonne, Illinois 60439, USA.
\end{affiliations}

\newpage

\begin{abstract}
On a lattice composed of triangular plaquettes where antiferromagnetic exchange interactions between localized spins cannot be simultaneously satisfied, the system becomes geometrically frustrated with magnetically disordered phases remarkably different from a simple paramagnet. Spin liquid belongs to one of these exotic states, in which a macroscopic degeneracy of the ground state gives rise to the rich spectrum of collective phenomena. Here, we report on the discovery of a new magnetic state in the heterostructures derived from a single unit cell  (111)-oriented spinel CoCr$_2$O$_4$ sandwiched between nonmagnetic Al$_2$O$_3$ spacers. The artificial quasi-two-dimensional material composed of three triangle and one kagome atomic planes shows a degree of magnetic frustration which is almost two orders of magnitude enlarged compared to the bulk crystals. Combined resonant X-ray absorption and torque magnetometry measurements confirm that the designer system exhibits no sign of spin ordering down to 30 mK, implying a possible realization of a quantum spin liquid state in the two dimensional limit.
\end{abstract}

\clearpage

\begin{figure}[htp]
\centering
\includegraphics[width=0.95\textwidth]{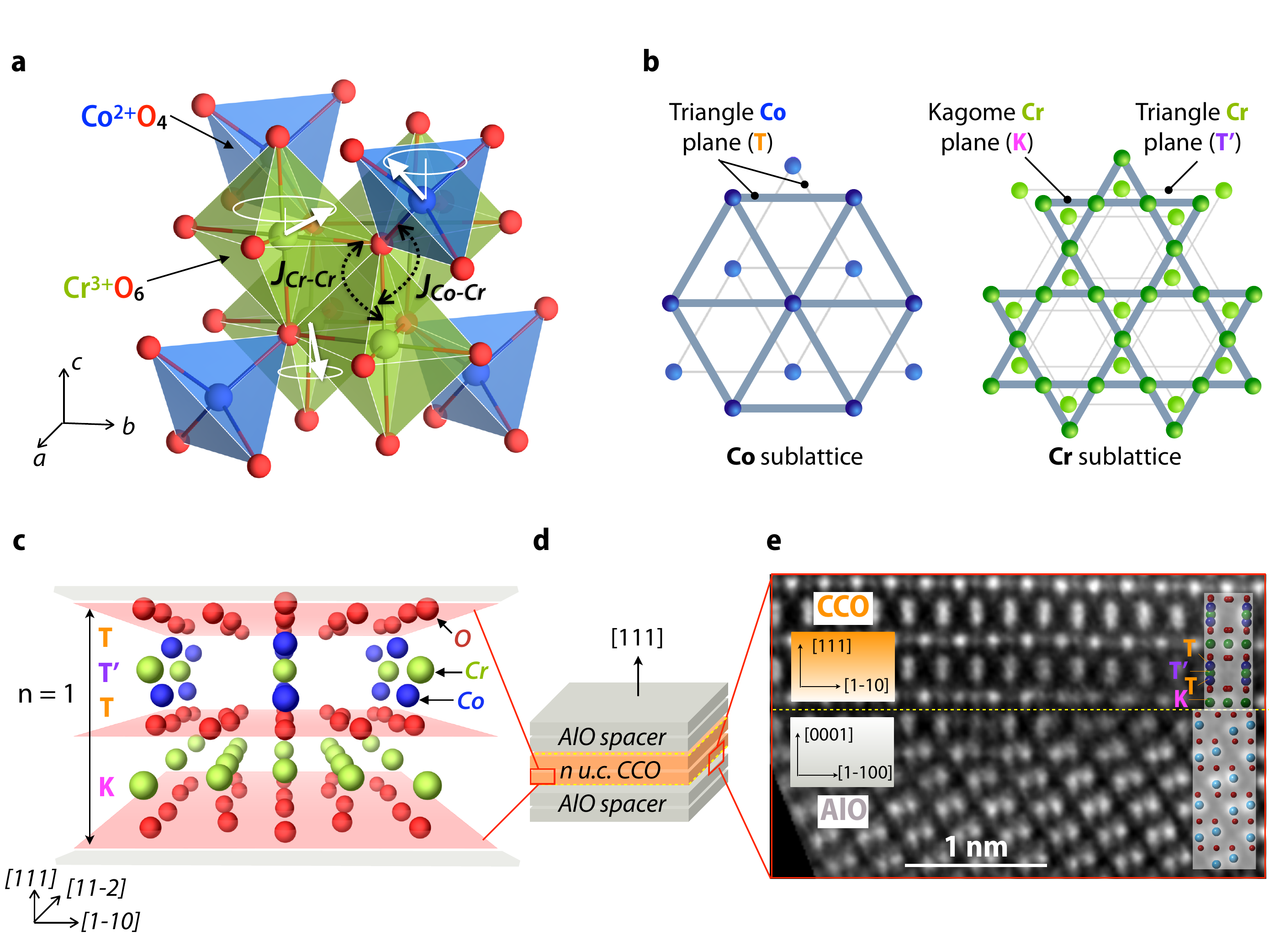}
\caption{\label{Fig1} {\bf Overview of the structural properties of the superlattices.} \textbf{a,} Schematic representation of bulk spinel CoCr$_2$O$_4$ composed of networks of Co$^{2+}$O$_4$ tetrahedra and Cr$^{3+}$O$_6$ octahedra. Both the inter-atomic and the intra-atomic superexchange pathways, $J_{\textrm{Cr-Co}}$ and $J_{\textrm{Cr-Cr}}$, are displayed on the graph with the conical spin structures (white arrows) as depicted by Yamasaki {\it{et al.}} \cite{Yamasaki_PRL_2006}. \textbf{b,} Stacking of the cation planes along the [111] direction. The blue Co ions form only triangle T plane (in orange), whereas the green Cr ions form both triangle T' plane (in purple) and kagome K plane (in pink). Oxygen ions are omitted for clarity. \textbf{c,} Definition of the basic repeating unit (\textit{n} = 1) of CoCr$_2$O$_4$ along the [111] direction, which includes four geometrically frustrated planes: kagome Cr plane (K), triangle Co plane (T), triangle Cr plane (T'), and triangle Co plane (T). \textbf{d,} Schematic illustration of the (111)-oriented (CoCr$_2$O$_4$)$_n$/(Al$_2$O$_3$)$_2$ superlattices.  \textbf{e,}
High-angle annular dark-field scanning transmission electron micrograph of the CoCr$_2$O$_4$/Al$_2$O$_3$ interface. Left inset: The epitaxial relationship
between two components. Right inset: Sketch of the relative positions of each type of ion (Co, Cr, Al, and O are depicted by dark blue, green, light
blue, and red spheres, respectively). The four cation planes included in one repeating unit are labeled on the image.}
\end{figure}

{\noindent In a magnetic crystal the basic notion of minimizing free energy necessitates that on cooling all spins should lock into a long-range ordered pattern. In sharp deviation from this, the combination of electronic correlations, quantum fluctuations, and lattice geometries supporting frustrated magnetic interactions has been a remarkably fertile ground for predicting unusual entangled states of quantum matter including spin ice and magnetic monopole, topological superconductors, axion insulators, Weyl semimetals, and a variety of liquid-like quantum spin phases \cite{Ramirez_review_1994,Balents_ARCMP_2014,Nisoli_RMF_2013}. For the latter, an entangled quantum spin liquid (QSL) state possesses no long range magnetic order, lacks any spontaneously broken symmetry, and carries a spectrum of fractional excitations \cite{Lee_Science_2008,Misguich_chap16_2010,Zhou_RMP_2017}. As for the experimental realization, to date the QSL state remains elusive mainly due to the ambiguity of response to local probes applied to the massively entangled many-body phase, the absence of a definitive experimental signature, and severe scarcity of bulk crystals to closely match the theoretical proposals \cite{Balents_nature_2010,Savary_IOP_2017}.} 

In recent years, however, complementary to bulk synthesis, ultra-thin heterostructures composed of two or more structurally, chemically and electronically dissimilar layers have been developed into a powerful platform for materials design and discovery \cite{Hwang_NatMater_2012,Chakhalian_RMP_2014}. Within this framework, in close synergy to interface and heteroepitaxial strain, geometrical lattice engineering (GLE) can be applied to forge unusual artificial lattice geometries by stacking a specific number of atomic planes along high-index crystallographic directions \cite{XL_MRS_2016,Sri_PRL_2016,Eom_Nature_2016}. Here, we exploit the GLE approach, to devise a new class of quasi-two-dimensional geometrically frustrated lattices derived from a single unit cell of the CoCr$_2$O$_4$ compound.\\    

{\noindent \bf Results}\\
{\bf Structures of the superlattices.}
CoCr$_2$O$_4$ belongs to the normal spinel (AB$_2$O$_4$) chromite family, MCr$_2$O$_4$ (M = Mn, Fe, Co and Ni), in which the magnetically active M$^{2+}$ ions occupy the tetrahedral A sites of diamond sublattice and the Cr$^{3+}$ ions occupy the octahedral B sites of pyrochlore sublattice \cite{Mufti_JPCM_2010}. Due to the competing effect of the antiferromagnetic nearest-neighbor exchange interactions between the inter-sublattices \textit{J}$_{\textrm{AB}}$ and the intra-sublattice \textit{J}$_{\textrm{BB}}$, the materials family is magnetically frustrated and exhibits complex magnetic ground states \cite{LKDM_PR_1962}. Specifically, in bulk CoCr$_2$O$_4$, a collinear ferrimagnetic state first forms with a Curie temperature of 93 K and further transforms into a spiral ferrimagnetic state with T$_S$ $\sim$ 26 K, as illustrated in Fig.~1A. Interestingly, when viewed along the [111] direction one unit cell of the same structure contains four geometrically frustrated atomic planes from the two sub-lattices giving rise to the ``Kagome Cr (K) -- Triangle Co (T) --Triangle Cr (T$^{\prime}$) -- Triangle Co (T)'' stacking sequence (see Figs.~1B and 1C). Based on this design idea a series of (111)-oriented [(CoCr$_2$O$_4$)$_n$/(Al$_2$O$_3$)$_2$]$_N$ superlattices ($\textit{n}$ = 1, 2 and 4 unit cells and  $\textit{n} \times N$ = 16) were fabricated by laser MBE on (0001)-oriented single crystal sapphire (alumina) substrate. For the multi-layer structure, $\alpha$-Al$_2$O$_3$ was selected as the non-magnetic spacer because of the excellent structural compatibility with CoCr$_2$O$_4$. Thickness of the active CoCr$_2$O$_4$ layer was varied from \textit{n} = 4 down to the two-dimensional limit of $\sim$ 5 \AA~(\textit{n} = 1), while thickness of the Al$_2$O$_3$ layer was kept constant at $\sim$ 10 \AA ~for all the samples (Fig.~1D). The high structural quality was confirmed by scanning transmission electron microscopy (Fig.~1E) demonstrating the presence of sharp interfaces with epitaxial registry CoCr$_2$O$_4$ (111) [1$\bar{1}$0]~$\|$~Al$_2$O$_3$ (0001) [1$\bar{1}$00]. Additionally, the issues of cation distribution disorder and valency mixture were eliminated by resonant X-ray absorption spectroscopy (XAS) measurements (Supplementary Fig.~1).\\

\begin{figure}[htp]
\centering
\includegraphics[width=0.95\textwidth]{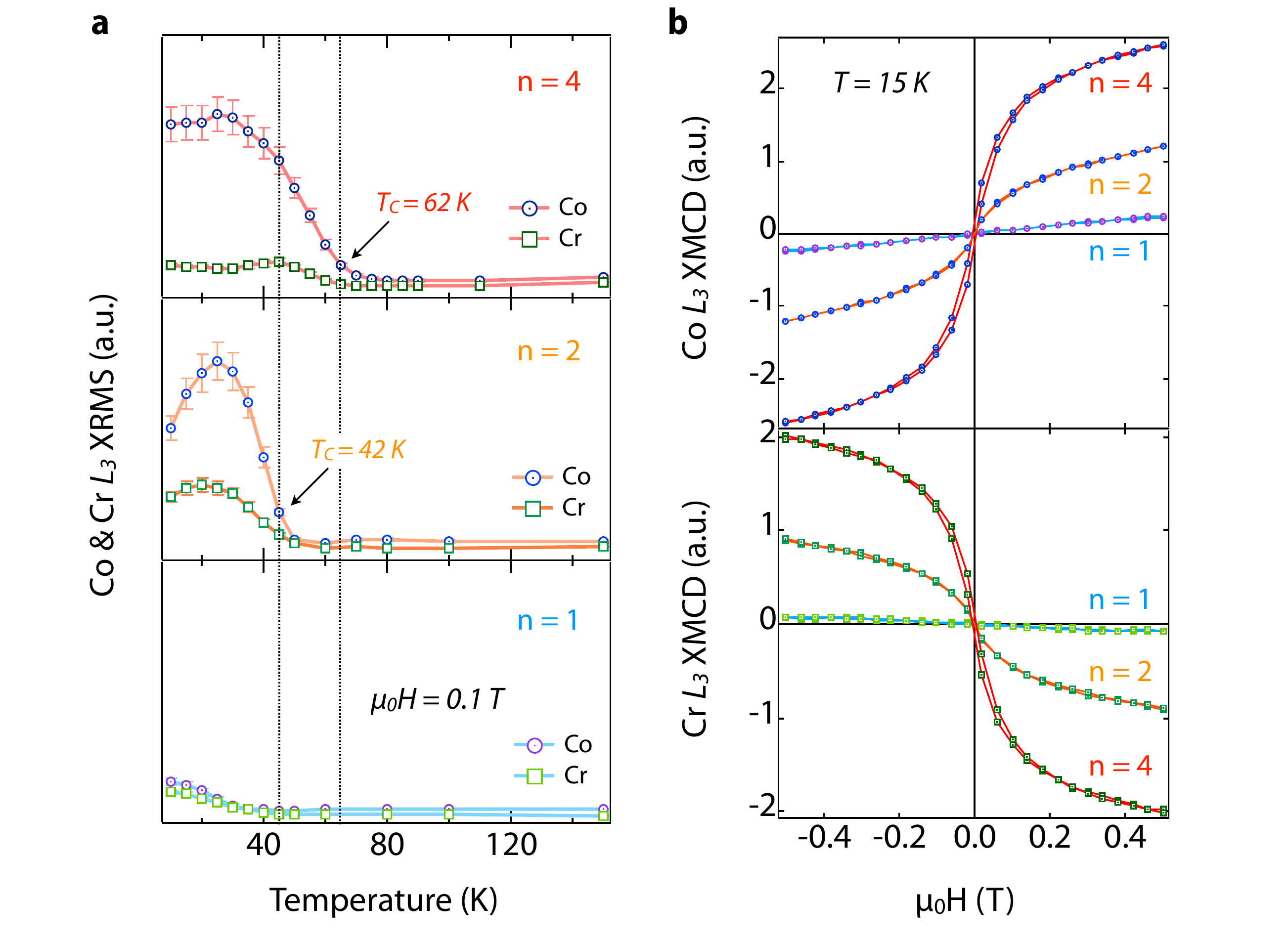}
\caption{\label{Fig2} {\bf Magnetic properties of the superlattices.} \textbf{a,} Temperature dependence of the XRMS $L_3$ peak intensity of both Co (circle) and Cr (square) for all the superlattices. Measurements were recorded using the reflectivity detection mode during the warming up process, after field-cooling the samples in a 0.1 T field from 300 K to 10 K. The onset temperature of the para- to ferri- magnetic transition is $\sim$ 62 K for \textit{n} = 4 and $\sim$ 42 K for \textit{n} = 2, whereas no distinct magnetic phase transition is observed on \textit{n} = 1. \textbf{b,} Field dependence of Co and Cr XMCD L$_3$ intensity, respectively, for all the superlattices. The scans were
performed by sweeping the external magnetic field up to 0.5 T after zero-field-cooling
from room temperature to 15 K.}
\end{figure}

{\noindent \bf Magnetic properties of the superlattices.}\
Next, we have investigated the magnetic properties of the superlattices by recording the resonant absorption spectra taken with left- and right-circularly polarized X-rays. The difference between those two spectra, or X-ray magnetic circular dichroism (XMCD) originates from the net magnetization of a specifically probed chemical element. The temperature dependence of the $L_3$-edge peak intensity shown in Fig.~2A confirms that for \textit{n} = 4 and \textit{n} = 2 samples, an onset of the magnetic phase transition occurs at T$_c$ $\sim$ 62 K and 42 K, respectively; Below T$_c$, the ferrimagnetic long-range order develops simultaneously on both Co and Cr sub-lattices. Surprisingly, \textit{n} = 1 sample shows no distinct signature of magnetic ordering and exhibits only a very moderate XMCD response at low temperatures in the presence of an applied magnetic field of 0.1 T. To further confirm the absence of spin ordering, XMCD hysteresis loops for both Co (at $\sim$ 778 eV) and Cr (at $\sim$ 577 eV) were measured on all samples at 15 K. As shown in Fig.~2B, the difference in the sign of the $L_3$ peak indicates the magnetic moments on Co and Cr sub-lattices are antiparallel (i.e. \textit{J}$_{\textrm{Co-Cr}} < 0$). In addition, the narrow loops detected for \textit{n} = 4 and \textit{n} = 2 samples imply that within each sub-lattice the spin moments are ferromagnetically aligned; In sharp contrast, for \textit{n} = 1 no hysteretic behavior is found and instead a linear XMCD vs. \textit{H} relationship typical of a magnetically disordered state appears on both Co and Cr ions. This observation provides another evidence for the absence of a long-range ordering down to 15 K. 



\begin{figure}[htp]
\centering
\includegraphics[width=0.9\textwidth]{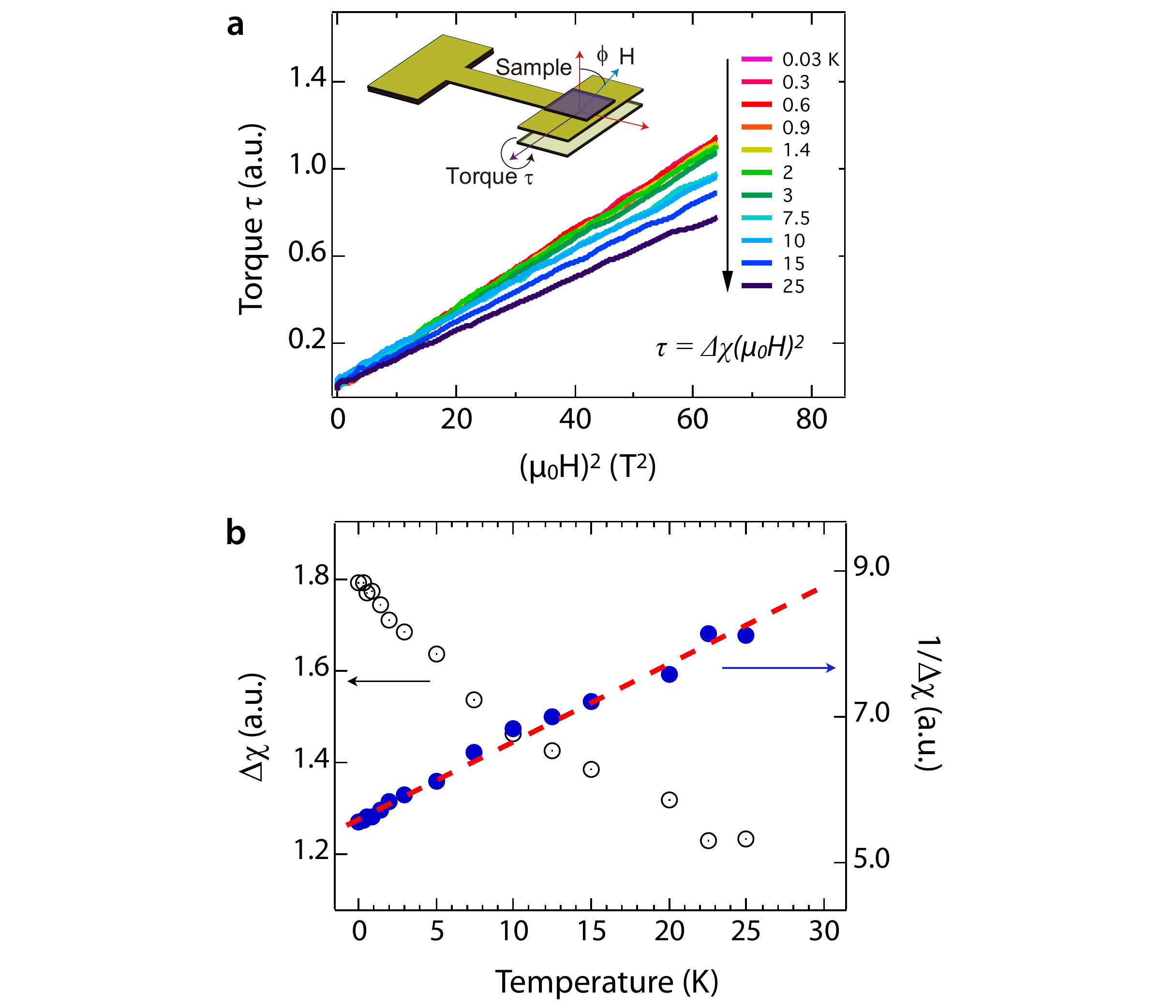}
\caption{\label{Fig3} {\bf Identification of the magnetic ground states for \textit{n} = 1.} \textbf{a,} Torque magnetometry curves at different temperatures from 25 K down to 0.03 K. The magnetic field was applied with $\phi$ = 30$^\circ$ away from the [111] direction, which is the surface normal of the film. Instead of showing any hysteresis loop behavior, all curves exhibit a $\tau \sim (\mu_0H)^2$ parabolic relationship in the entire temperature range. \textbf{b,} Calculated $\Delta \chi$ (left scale) and $1 / \Delta \chi$ (right scale) with linear fit (red dashed line) as a function of temperature.}
\end{figure}

For \textit{n} = 1, in order to examine if any long-range magnetic order emerges at extremely low temperature, we performed torque magnetometry measurements from 30 K down to 0.03 K. This technique quantifies the magnetic torque response of a sample with respect to the applied field ($\bf{\tau} = \mu_0 \bf{M} \times \bf{H}$), and has been demonstrated a powerful utility to probe extremely small magnetic signals from ultrathin samples and interfaces \cite{Li_Science_2008,Li_NatPhys_2011}. As clearly seen in Fig.~3A, within resolution of the measurement and down to the base temperature, no hysteresis is observed and instead a reversible parabolic $\tau \sim (\mu_0H)^2$ relationship is found, implying that the \textit{n} = 1 sample has neither long range ordering nor magnetic domain formation. Moreover, the effective susceptibility $\Delta \chi$ obtained from the torque data exhibits a smooth variation as temperature drops from 30 K to 0.03 K, excluding the presence of any magnetic phase transition.\\

{\noindent \bf Identification of the ground state.}\
In the absence of an obvious magnetic phase transition, thus far a crucial question remains: What is the magnetic ground state for \textit{n} = 1? Associating with the effect of magnetic frustration, to answer this question, we did the Brillouin function fittings on the XMCD vs. $H$ curves for the complete set of samples (see Supplementary Information). Table~1 summarizes the observed transition temperature T$_c$, along with the obtained values of $\Theta_{\textrm{cw}}$, and the corresponding frustration factors $f = T_c / \Theta_{\textrm{cw}}$ as a function of thickness \textit{n}. As immediately seen, compared to the bulk value \cite{Tomiyasu_PRB_2004}  $f (\textit{n} = \infty) \sim$ 6, moving down to \textit{n} = 2 the degree of frustration becomes monotonically suppressed on a moderate scale, followed by the dramatic jump by over two orders of magnitude for \textit{n} = 1. These findings signal that as the layer thickness reaches the two-dimensional limit, the magnetic behavior of (111)-oriented CoCr$_2$O$_4$ enters into an entirely different from the bulk regime. 

\begin{table}
\caption{{\bf Evolution of the magnetic frustration in the superlattices.} The fitted Curie-Weiss temperature $\Theta_{cw}$, the observed magnetic phase transition temperature T$_c$, and the calculated frustration factor $f$ are shown for both the CoCr$_2$O$_4$ bulk and the superlattices with \textit{n} = 4, 2, and 1.}\

\centering
\setlength{\tabcolsep}{4.5pt}
\begin{tabular}{c c c c}
\hline\hline
Sample & $\Theta_{cw}$ (K) & T$_c$ (K) & $f$ \\ [0.5ex]
\hline
Bulk & 582 & 97 & 6.0 \\
n = 4 & 92 & 62 & 1.5 \\
n = 2 & 57 & 42 & 1.4 \\
n = 1 & 4.96$\pm$0.94 & $<$ 0.03 & $>$ 165.3$\pm$31.3 \\
\hline\hline
\end{tabular}
\label{Table I}
\end{table}

In modern theory of frustrated magnetism, the absence of magnetic ordering down to the lowest accessible temperature coupled with a large value of $f \sim 100 - 1000$ serves as the key `smoking gun' for a quantum spin liquid (QSL) \cite{Ramirez_review_1994, Zhou_RMP_2017, Balents_nature_2010, Savary_IOP_2017,Pratt_Nature_2011}. In addition, a further evidence typically comes from a spectrum of magnetic excitations delivered by neutron scattering \cite{Nakatsuji_science_2005,Ross_PRX_2011,Hirschberger_science_2015}. 
However, when applied to the artificial heterostructures with fewer unit cells, this task falls into a category of yet unresolved experimental problems. Nevertheless, based on the obtained results, it would be interesting to speculate if the magnetically disordered state for \textit{n} = 1 can be one of the candidates. First, we note that theory treats QSL as a distinctly entangled quantum phase beyond the conventional Ginzburg-Landau paradigm, which in turn presents an experimental challenge in identifying a QSL by the majority of local probes \cite{Savary_IOP_2017}. The presence of QSL, however, can be conjectured by excluding other competing ground states. First, it is clearly not a simple paramagnet due to the opposite sign of the XMCD loops (otherwise, all the spins on both Co and Cr sites should cant along the external field). As for other spin-disordered phases such as spin glass, one expects a ``cusp'' to develop in the $M$ vs. $T$ below a characteristic temperature $T_g$ \cite{Binder_review_1986}. Again, no such ``cusp'' behavior is observed in the temperature-dependent torque and XMCD measurements for \textit{n} = 1. Last, judging by the absence of any spin-flip (or spin-flop) feature up to 8 T, and the rather smooth temperature-dependent $\Delta\chi$ with no anomaly, we can rule out the ground state of \textit{n} = 1 with long-range antiferromagnetic ordering.\\ 

\begin{figure}[htp]
\includegraphics[width=\textwidth]{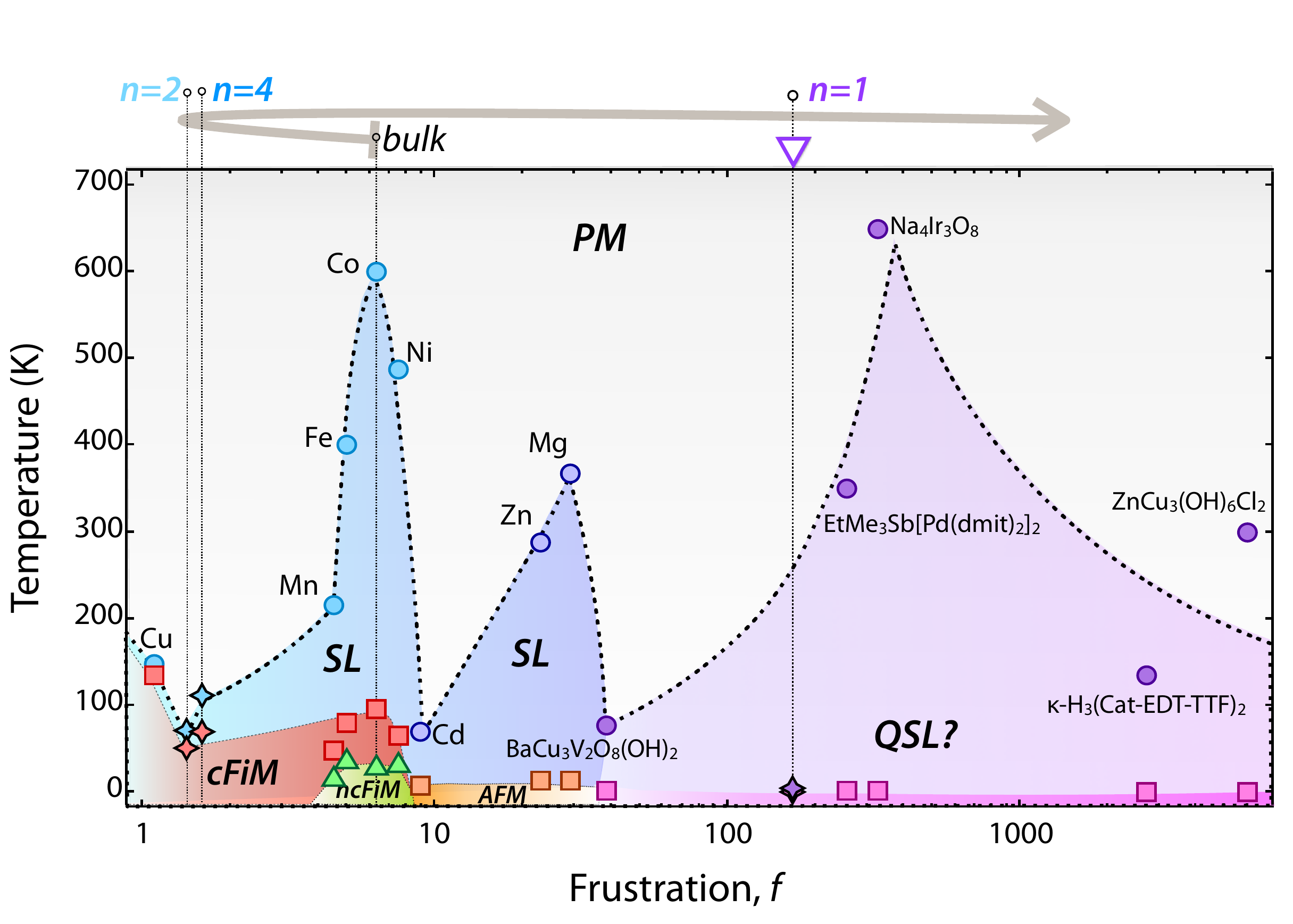}
\caption{\label{Fig4} {\bf Phase diagram as a function
of frustration parameter.} Definition of the symbols: PM --- paramagnet,
SL --- spin liquid, cFiM --- collinear ferrimagnet, ncFiM --- non-collinear
ferrimagnet, AFM --- antiferromagnet, QSL --- quantum spin liquid. The circle points crossed by the dotted line represent the fitted Curie-Weiss temperature $\Theta_{cw}$, while the square and the triangular points represent the real magnetic transition temperature. Data for MCr$_2$O$_4$ with different A sites are taken from
literatures (M = Cu\cite{Suchomel_PRB_2012}, Mn\cite{Dey_PRB_2014}, Fe\cite{Shirane_JAP_1964},
Co\cite{Tsurkan_PRL_2013}, Ni\cite{Suchomel_PRB_2012};
Cd, Zn and Mg\cite{Kemei_JPCM_2013}).
Data for the superlattices with \textit{n} = 1, 2 and 4 are displayed in diamond.
The gray arrow on the top represents the evolution of the frustration effect
in (111)-oriented CoCr$_2$O$_4$ towards the two-dimensional limit. Experimental results
of possible QSL candidates with various types of lattice are plotted for
comparison (EtMe$_3$Sb[Pd(admit)$_2$]$_2$\cite{Itou_PRB_2008} and $\kappa$
-- H$_3$(Cat-EDT-TTF)$_2$\cite{Isono_PRL_2014} with triangle lattice; Na$_4$Ir$_3$O$_8$\cite{Okamoto_PRL_2007}
with hyperkagome lattice; BaCu$_3$V$_2$O$_8$(OH)$_2$\cite{Okamoto_JPSJ_2009}
and ZnCu$_3$(OH)$_6$Cl$_2$\cite{Helton_PRL_2007} with kagome lattice). In this regime, the pink square points represent the lowest temperatures ever reached in literature, where the systems still have no magnetic orderings.}
\end{figure}

{\noindent \bf Evolution of the magnetic frustration.}\
As illustrated in Fig.~4, to put our results in perspective with respect to the several selected cases of disordered quantum magnetism, we establish the ``temperature -- frustration'' phase diagram including our samples, members of  ACr$_2$O$_4$ family \cite{Suchomel_PRB_2012,Dey_PRB_2014,Shirane_JAP_1964,Tsurkan_PRL_2013,Kemei_JPCM_2013} and well recognized candidates for QSL\cite{Itou_PRB_2008,Isono_PRL_2014,Okamoto_PRL_2007,Okamoto_JPSJ_2009,Helton_PRL_2007}. First, consider the magnetic behavior of the spinel family. When the A site is magnetic (e.g. A = Cu, Mn, Fe, Co and Ni), the dominant inter-sublattice antiferromagnetic super-exchange coupling $J_{\textrm{AB}}$ favors the N\'{e}el-type ferrimagnetic configuration but still competes with the antiferromagnetic interaction, $J_{\textrm{BB}}$ of the B-site pyrochlore sublatttice, leading to a weak geometrical frustration; Depending on the degree of frustration, the magnetic ground state can be either collinear or non-collinear ferrimagnetic.\cite{Tomiyasu_PRB_2004,LKDM_PR_1962}
On the other hand, when the A site is non-magnetic (e.g. A = Cd, Zn and Mg) the antiferro-magnetically coupled Cr moments of the pyrochlore sub-lattice are subjected to the much stronger geometrical frustration thus maintaining the spin liquid state over a wider temperature range. If quantum fluctuations are not sufficiently strong, a long-range antiferromagnetic order with a complex spin configuration can still emerge. Potentially, further increase of quantum fluctuations may drive the system into the region of extremely high frustration, where the disordered spin state persists down to 0 K. Very recently, this scenario \cite{Mila_EJP_2000} was realized on the lattices with reduced dimensionality and coordination (e.g. BaCu$_3$V$_2$O$_8$(OH)$_2$ to ZnCu$_3$(OH)$_6$Cl$_2$ in Fig.~4).

In our heterostructures the effect of reduced dimensionality is to localize electrons through the enhancement of electron-electron correlations, $U/W$. Theoretically, in CoCr$_2$O$_4$ the increased value of $U/W$ leads to the reduction in strength of all exchange constants, which in turn reduces the Curie-Weiss temperature $\Theta_{\textrm{cw}}$  \cite{Ederer_PRB_2007}. Moreover, the theory predicts that the decrease of $J_{\textrm{Cr-Cr}}$ occurs faster than $J_{\textrm{Co-Cr}}$\cite{Ederer_PRB_2007}, which consequently suppresses the effect of competing interactions and lowers the degree of magnetic frustration. The experimental results for \textit{n} = 4 and \textit{n} = 2 samples are in a good agreement with this picture, where we observe reduced values of both $\Theta_{\textrm{cw}}$ and $f$. In particular, as shown in Fig.~4, \textit{n} = 4 and \textit{n} = 2 samples with a collinear ferrimagnetic ground state are indeed shifted to the left. In sharp constrast, the extremely large value of $f$  for \textit{n} = 1 indicates that this system bypasses the entire ACr$_2$O$_4$ region with no bulk-like imprint. Such a marked increase of $f$ may originate from several important contributions. First, as the \textit{n} = 1 heterostructure has only four geometrically frustrated cation planes, it approaches the two-dimensional limit where quantum fluctuations are naturally expected to be strongly enhanced. Secondly, because of breaking of the translational symmetry along the [111] direction, the in-plane antiferromagnetic exchange interactions on those four geometrically frustrated lattices prevail and result in a drastically enlarged value of $f$. 

To summarize, we have provided detailed microscopic insights into the nature of the magnetic ground state of (111)-oriented (CoCr$_2$O$_4$)$_n$/(Al$_2$O$_3$)$_2$ superlattices and discovered the emergence of a novel magnetic state for the single unit cell thin heterostructure. In this ultimate ultra-thin limit (\textit{n} = 1), the drastically increased value of magnetic frustration and the absence of spontaneous magnetic ordering down to 30 mK reveal an unusual shift from the modified bulk-like behavior to the new disordered quantum phase with a strong potential for a QSL state. These findings illustrate the power of geometrical lattice engineering to realize exotic many-body phenomena and may conceivably put ``an end to the drought of quantum spin liquids''. \cite{Lee_Science_2008} In addition to the technological importance of artificial planar geometries with QSL behavior for quantum information processing \cite{Wilczek_NP_2009}, the availability of such heterostructures opens the opportunity for electron doping by gating to reach enigmatic topological superconductivity \cite{Kelly_PRX_2016}.\\


\begin{methods}

\noindent{\bf {Sample fabrication.}}\ The superlattices were grown on 5 $\times$ 5 mm$^2$ Al$_2$O$_3$ (0001) substrates by pulsed laser deposition. Stoichiometric CoCr$_2$O$_4$ and Al$_2$O$_3$ targets were ablated using a KrF excimer laser ($\lambda$ = 248 nm, energy density $\sim$ 2 Jcm$^{-2}$) with a repetition rate of 4 Hz and 2 Hz, respectively. The depositions were carried out at a substrate temperature of 700 $^{\circ}$C, under oxygen partial pressure of 5 mTorr. The films were post-annealed at the growth condition for 15 minutes and then cooled down to room temperature. The layer-by-layer growth was monitored by {\it in-situ} high pressure reflection-high-energy-electron-diffraction (RHEED) during the deposition process. Furthermore, quality of the superlattices has been verified by using a variety of {\it ex-situ} characterization methods including synchrotron-based X-ray diffraction, X-ray reflectivity, X-ray photoemission spectroscopy and atomic force microscopy, as described in detail elsewhere \cite{Xiaoran_APL_2014, Xiaoran_APL_2015}.\

\noindent{\bf {Scanning transmission electron microscopy.}}\ The scanning transmission electron microscopy (STEM) measurements were carried out using a spherical aberration-corrected JEM-ARM200F, operated at 200 kV. The high-angle annular dark-field (HAADF) imaging was taken using the collection semi-angle of about 70-250 mrad.\ 

\noindent{\bf {DC magnetization measurements.}}\ The magnetization of the superlattices were measured using the superconducting quantum interference devices (SQUID) from quantum design (see Supplementary Figure 2). The signals were recorded during the zero field cooling process from 300 K down to 2 K. The measurements were repeated 8 times on each sample to gain sufficient statistics. In order to distinguish the magnetic transition temperature and elucidate any artifacts, the temperature-dependent magnetization curve was also recorded for a bare Al$_2$O$_3$ substrate, which was treated with the same thermal process (heating, annealing and cooling) as growing the films.\       

\noindent{\bf {Synchrotron X-ray absorption spectroscopy.}}\ The X-ray absorption spectroscopy (XAS) experiments were performed at beamline 4-ID-C of the Advanced Photon Source at Argonne National Laboratory. The L-edge X-ray absorption spectra of both Co and Cr were scanned with left- and right- polarized X-rays at grazing incidence to obtain the X-ray magnetic circular dichroism (XMCD). Samples were first field cooled down to 10 K and then measured during the warming up process. To exclude any artifact, all the measurements were conducted in both positive and negative external field. Data were recorded simultaneously with total electron yield (TEY), fluorescence yield (FY), and X-ray magnetic reflectivity (XRMS) detection modes. In addition, the XMCD hysteresis loops were measured at beamline 4.0.2 of the Advanced Light Source at Lawrence Berkeley National Laboratory. Samples were cooled with zero field and maintained at 15 K. The circularly polarized soft X-rays were incident with an angle of 35$^\circ$ relative to the sample surface. The luminescence detection mode was used to record the data.\

\noindent{\bf {Torque magnetometry measurements.}}\ Torque magnetometry measurements were performed with our home built cantilever setup by attaching samples to a thin beryllium copper cantilever. External magnetic field was applied, and the sample rotation by torque was measured by tracking the capacitance change between the metallic cantilever and a fixed gold film underneath using an AH2700A capacitance bridge with a 14 kHz driving frequency. To calibrate the spring constant of the cantilever, we tracked the angular dependence of capacitance caused by the sample weight at zero magnetic field.\


\end{methods}

\begin{addendum}
 \item [Acknowledgements.]
The authors deeply acknowledge D. Khomskii, G. Fiete, X. Hu, P. Coleman, K. Rabe, P. Chandra and P. Mahadevan for numerous insightful discussions. X.L. and J.C. acknowledge the support by the Gordon and Betty Moore Foundation EPiQS Initiative through Grant No. GBMF4534, and B.P. by the Department of Energy under Grant No. DE-SC0012375. The research of L.L. is supported by the US Department of Energy, Office of Basic Energy Sciences, Division of Materials Sciences and Engineering under Award DE-SC0008110 (high-field magnetization). Q.Z. and L.G. are supported by the Strategic Priority Research Program of the Chinese Academy of Science, Grant No. XDB07030200. This research used resources of the Advanced Light Source, which is a Department of Energy Office of Science User Facility under Contract No. DE-AC0205CH11231. This research used resources of the Advanced Photon Source, a U.S. Department of Energy Office of Science User Facility operated by Argonne National Laboratory under Contract No. DE-AC02-06CH11357. We thank these institutions for continuing assistance.
 
 

\end{addendum}





\end{document}